\documentclass[pre,twocolumn,showpacs,superscriptaddress]{revtex4-1}
\pdfoutput=1
\usepackage{amsmath,amssymb}
\usepackage[xdvi]{graphicx}%
\usepackage{dcolumn}
\usepackage{amsmath}
\usepackage{bm}
\usepackage{subfigure}

\usepackage[utf8]{inputenc}
\usepackage[T1]{fontenc}
\usepackage{ae,aecompl}

\usepackage{color}
\definecolor{darkblue}{rgb}{0,0,0.6}
\definecolor{darkred}{rgb}{0.6,0,0}
\definecolor{darkgreen}{rgb}{0,0.6,0}

\usepackage[colorlinks=true,urlcolor=darkblue,citecolor=darkblue,linkcolor=darkred,hyperfootnotes=false]{hyperref}

\newcommand{\dd}{\mathrm{d}}

\newcommand{\cc}{\mathrm{c}}

\begin{document}

\title{Dynamical symmetry breaking and phase transitions \\ in driven diffusive systems}

\author{Yongjoo Baek}
\email{yongjoo.baek@physics.technion.ac.il}
\affiliation{Department of Physics, Technion, Haifa 32000, Israel}

\author{Yariv Kafri}
\affiliation{Department of Physics, Technion, Haifa 32000, Israel}

\author{Vivien Lecomte}
\affiliation{Laboratoire Probabilit\'{e}s et Mod\`{e}les Al\'{e}atoires, UMR7599 CNRS, Sorbonne Paris Cit\'e, Universit\'{e} Pierre et Marie Curie \& Universit\'{e} Paris Diderot, F-75013 Paris, France}

\date{\today}

\begin{abstract}

We study the probability distribution of a current flowing through a diffusive system connected to a pair of reservoirs at its two ends. Sufficient conditions for the occurrence of a host of possible phase transitions both in and out of equilibrium are derived. These transitions manifest themselves as singularities in the large deviation function, resulting in enhanced current fluctuations. Microscopic models which implement each of the scenarios are presented, with possible experimental realizations. Depending on the model, the singularity is associated either with a particle--hole symmetry breaking, which leads to a continuous transition, or in the absence of the symmetry with a first-order phase transition. An exact Landau theory which captures the different singular behaviors is derived.

\end{abstract}

\maketitle

In recent years there has been much activity focused on understanding probability distributions in systems which are far from thermal equilibrium. In particular, the probability of observing a current flowing between two reservoirs, through an interacting channel, was studied in many works for both quantum \cite{levitov1993pis,levitov_electron_1996,pilgram_stochastic_2003,jordan_fluctuation_2004,flindt_trajectory_2013} (in the context of `full counting statistics') and classical systems~\cite{derrida_exact_1998,derrida_universal_1999,derrida_current_2004,bodineau_current_2004,bertini_current_2005,bodineau_distribution_2005,bertini_non_2006,Bodineau:2007iq,prolhac_current_2008,appert-rolland_universal_2008,Imparato2009,prolhac_cumulants_2009,baek_large_n_2016,hurtado_test_2009,prados_large_2011,de_gier_large_2011,lazarescu_exact_2011,hurtado_spontaneous_2011,derrida_microscopic_2011,gorissen_current_2012,gorissen_exact_2012,krapivsky_fluctuations_2012,meerson_extreme_2013,meerson_extreme_2014,znidaric_exact_2014,Hurtado:2014bn,Lazarescu2015,Shpielberg2016,Zarfaty:2016dv,hirschberg_zrp_2015}. The properties of the distribution encode much information about the interactions in the channel. 

One of the most dramatic consequences of such interactions is the occurrence of dynamical phase transitions (DPT)~\cite{derrida_lebowitz_speer_2002,hirschberg_zrp_2015,bertini_current_2005,bodineau_distribution_2005,Bodineau:2007iq,lecomte_thermodynamic_2007,garrahan_dynamical_2007,bunin_non-differentiable_2012,baek_singularities_2015,tizon-escamilla_order_2016}, which are the focus of this Letter~\footnote{We also note that there are DPTs associated with changes of mean behaviors as boundary conditions are varied, most notably those of the ASEP (see \cite{Blythe2007} for a review). These are phenomena of a very different origin.}. They imply an enhanced probability of observing certain current fluctuations. Beyond certain current thresholds, the mode of transport through the channel changes abruptly. These DPTs manifest themselves as singularities in a \emph{large deviation function} (LDF) that characterizes the probability distribution of the time-averaged current $J$ in the limit of a large observation time. The function plays, for time-integrated observables, like $J$, the same role as the equilibrium free energy for static observables~\cite{touchette_large_2009}. For classical interacting particles systems, it can be computed using exact microscopic solutions~\cite{derrida_exact_1998,derrida_universal_1999,derrida_current_2004,prolhac_current_2008,prolhac_cumulants_2009,lazarescu_exact_2011,de_gier_large_2011,derrida_microscopic_2011,gorissen_exact_2012,Lazarescu2015} or macroscopic approaches (see~\cite{bertini_macroscopic_2015} for a review).

So far, for current large deviations in driven diffusive systems, only one class of DPTs with concrete microscopic models has been observed; these occur solely for periodic systems which are not connected to reservoirs~\cite{bertini_current_2005,bodineau_distribution_2005,Bodineau:2007iq,appert-rolland_universal_2008,tizon-escamilla_order_2016}. There one finds that, for currents close to the mean value, the fluctuation manifests itself through a {\it time-independent} density profile. The DPT occurs at a critical value of the current beyond which the fluctuation is realized through a {\it time-dependent} density profile. Such transitions are referred to as resulting from a failure of the `additivity principle'~\cite{bodineau_current_2004}. Another scenario which involves a `first-order' transition between two distinct time-independent density profiles was suggested in~\cite{bertini_non_2006}. However, lacking any concrete microscopic model, the scenario remains speculative.

In this Letter we study current large deviations in one-dimensional diffusive systems coupled to two reservoirs. Based on an exact Landau theory for the DPTs derived using the Macroscopic Fluctuation Theory (MFT)~\cite{bertini_fluctuations_2001,*bertini_macroscopic_2002,*bertini_minimum_2004,*bertini_stochastic_2007,bertini_macroscopic_2015}, we obtain the following new results: first, we identify DPTs that are not associated with a breaking of the additivity principle, along with sufficient conditions for their existence in terms of transport coefficients; second, we describe a new type of `second-order' DPTs associated with a \emph{symmetry breaking} in the density profiles which realize the current fluctuations; third, we show that well-studied microscopic models, namely the Katz--Lebowitz--Spohn (KLS)~\cite{katz_nonequilibrium_1983} model and the weakly asymmetric simple exclusion process (WASEP)~\cite{WASEP,gartner_convergence_1987}, implement both the first- and the second-order DPTs described above; finally, possible experimental realizations are discussed.

{\em Settings} --- We consider a one-dimensional driven diffusive system connecting two particle reservoirs using the standard approach of fluctuating hydrodynamics~\cite{spohn_long_1983,Spohn_Book,jordan_fluctuation_2004,bertini_macroscopic_2015}. The particle density profile $\rho(x,t)$ evolves according to a continuity equation
\begin{align} \label{eq:continuity}
\partial_t \rho(x,t) + \partial_x j(x,t) = 0\,,
\end{align}
where the spatial coordinate $x$ is rescaled by the system size $L$ so that $x \in [0,1]$, $t$ denotes time measured in units of $L^2$, and $j(x,t)$ is the fluctuating current given by
\begin{align} \label{eq:current}
j(x,t) = -D(\rho)\partial_{x}\rho + \sigma(\rho)E + \sqrt{\sigma(\rho)}\eta(x,t) \, .
\end{align}
The current consists of contributions from Fick's law, the response to a bulk field $E$, and a noise term. The diffusivity $D(\rho)$ and the mobility $\sigma(\rho)$ are in general density-dependent and connected by the Einstein relation, $2D(\rho)/\sigma(\rho)=\partial_\rho^2 f(\rho)$, with $f(\rho)$ the free energy density of the system at equilibrium. The noise $\eta(x,t)$ satisfies $\langle \eta(x,t) \rangle = 0$ and
\begin{align} \label{eq:noise}
\langle \eta(x,t)\eta(x',t') \rangle = L^{-1}\delta(x-x')\delta(t-t') \,,
\end{align}
where $\langle \cdot \rangle$ denotes an average over all realizations of the noise. The spatial boundary conditions are fixed as $\rho(0,t) = \bar\rho_a$ and $\rho(1,t) = \bar\rho_b$, where $\bar\rho_a$ and $\bar\rho_b$ are time-independent densities imposed by the reservoirs. We are interested in phase transitions~\cite{bertini_current_2005,bodineau_distribution_2005,Bodineau:2007iq} associated with the time-averaged current
\begin{align} \label{eq:J_def}
J \equiv \frac{1}{T} \int_0^T \mathrm{d}t \int_0^1 \mathrm{d}x \, j(x,t)\,,
\end{align}
whose statistics obey a large deviation principle \cite{bertini_macroscopic_2015}
\begin{align} \label{eq:J_ldp}
P(J) \sim \exp \left[-TL\Phi(J)\right]
\end{align}
for $T \gg 1$. A singularity in the LDF $\Phi(J)$ marks a DPT. It proves to be convenient to change ensembles and work with the scaled cumulant generating function (CGF)
\begin{align} \label{eq:scgf_def}
\Psi(\lambda) \equiv \lim_{T \to \infty} \frac{1}{TL} \ln \left\langle e^{TL \lambda J} \right\rangle\,.
\end{align}
Standard saddle-point arguments~\cite{touchette_large_2009} show that the scaled CGF is related to the LDF by a Legendre transform $\Psi(\lambda) = \sup_J \left[\lambda J - \Phi(J) \right]$. To calculate $\Psi(\lambda)$, we rewrite Eq.~\eqref{eq:scgf_def} in a path integral form using the Martin--Siggia--Rose formalism~\cite{martin_statistical_1973,*dominicis_technics_1976,*janssen_lagrangean_1976}, which gives
\begin{align}
	\Psi(\lambda) = \lim_{T\to\infty}\frac{1}{TL} \ln \int \mathcal{D}\rho \mathcal{D}\hat\rho \, e^{-L\int_0^T \mathrm{d}t \int_0^1 \mathrm{d}x \, \left[\hat\rho\partial_t\rho - H(\rho,\hat\rho)\right]} \,,
\end{align}
with the Hamiltonian density $H(\rho,\hat\rho)$ defined as
\begin{align}
  \hspace*{-1.5mm}
H(\rho,\hat\rho) \equiv -D(\rho)(\partial_x\rho)(\partial_x\hat\rho) + \tfrac{\sigma(\rho)}{2}(\partial_x\hat\rho)(2E+\partial_x\hat\rho) .\!
\label{eq:hamiltonian}
\end{align}
The `momentum' variable $\hat\rho$ satisfies the boundary conditions~(see Appendix~\ref{sec:saddle_point_eqns}) $\hat\rho(0) = 0$ and $\hat\rho(1) = \lambda$. The scaled CGF $\Psi(\lambda)$ can then be obtained using a saddle-point method. For our cases of interest, we argue that the saddle-point solutions are time-independent, so that the additivity principle is satisfied. The calculations are detailed in Appendices~\ref{sec:eq_bcs} and \ref{sec:ueq_bcs}, and yield profiles $\rho^*(x)$ and $\hat\rho^*(x)$ which minimize the action $\int_0^T \mathrm{d}t \int_0^1 \mathrm{d}x \, \left[\hat\rho\partial_t\rho - H(\rho,\hat\rho)\right]$. These profiles, which are called the \emph{optimal profiles}, represent the dominant realizations of current fluctuations at a given value of $\lambda$. As we will see, phase transitions are associated with abrupt changes in the shape of the optimal profile as $\lambda$ is varied.

{\em Results} --- In what follows, we first consider systems with equal boundary densities $\bar\rho_a = \bar\rho_b = \bar\rho$ with $\bar\rho$ very close to an {\em extremum} of $\sigma(\rho)$. Already in this case, depending on $D(\rho)$ and $\sigma(\rho)$, all singular behaviors described above are observed. Interestingly, this includes systems which are {\em in equilibrium}. Then, for more general boundary conditions given by $\bar\rho_a = \bar\rho - \delta\rho$ and $\bar\rho_b = \bar\rho + \delta\rho$, we argue perturbatively to the leading order in $\delta\rho$ that the behaviors are unchanged up to a shift of the transition point.

As shown in Appendix~\ref{sec:eq_bcs}, the problem of minimizing over profiles can be reexpressed as
\begin{align} \label{eq:psi}
\Psi(\lambda) = \frac{\bar\sigma}{2}\,\lambda\,(\lambda+2E) - \inf_m \mathcal{L}(m) \,,
\end{align}
where the Landau-like function $\mathcal{L}(m)$ of the parameter $m\in \mathbb R$, which captures the singular behaviors of $\Psi(\lambda)$, can be written in a truncated form
\begin{align} \label{eq:landau_sym}
\mathcal{L}(m) &\simeq 
	-\frac{2\pi \bar D^2}{\bar\sigma\bar\sigma''} \, \bar\sigma' \, m
	-\frac{(\lambda_\cc + E) \bar\sigma''}{4} \,(\lambda - \lambda_\cc) \, m^2 \nonumber\\
	& \quad-\frac{2\pi \bar D(\bar D \bar\sigma^{(3)}-3\bar D'\bar\sigma'')}{9\bar\sigma\bar\sigma''}
	\,m^3 \nonumber\\
	&\quad + \left[\frac{\pi^2\bar D\left(4\bar D'' \bar\sigma'' - \bar D \bar\sigma^{(4)}\right)}{64\bar\sigma\bar\sigma''}
	+ \frac{\bar\sigma''^2 E^2}{64\bar\sigma}
	\right] m^4 \, .
\end{align}
Here $\lambda_\cc$ is equal to one of the two values~\footnote{This expression for $\lambda_\cc$ is consistent with \cite{Imparato2009}, in which a condition for $\lambda_\cc$ was derived for the special case of constant $D(\rho)$, quadratic $\sigma(\rho)$, and $E = 0$.}
\begin{align} \label{eq:lc_def}
\lambda_\cc^\pm \equiv -E \pm \sqrt{E^2 + \frac{2\pi^2 \bar D^2}{\bar\sigma \bar\sigma''}} \,,
\end{align}
and we use the shorthand notations $\bar g \equiv g(\bar\rho)$, $\bar g' \equiv g'(\bar\rho)$, $\bar g'' \equiv g''(\bar\rho)$, and $\bar g^{(n)} \equiv g^{(n)}(\bar\rho)$ for derivatives of any function $g(\rho)$ evaluated at $\rho = \bar\rho$. The optimal value of the order parameter $m$ in Eq.~\eqref{eq:psi}, which we denote by $m^*$, measures the deviation of the optimal profile from the flat reference profile of density $\bar\rho$ (similar to the zero magnetization in the Landau theory for the Ising model):
\begin{align}
\rho^*(x) = \bar\rho + m^* \sin (\pi x) + O\left[(m^*)^2\right] \, .
\label{eq:defrhostar} 
\end{align}
The scaled CGF $\Psi(\lambda)$ has a singularity when $m^*$ changes in a singular manner as $\lambda$ is varied~\footnote{We note that a similar Landau theory for DPTs in periodic systems has been constructed in \cite{Bodineau:2007iq}.}.

Clearly, $\mathcal{L}(m)$ can be truncated as in Eq.~\eqref{eq:landau_sym} only if the coefficient of $m^4$ is positive. For the microscopic models we study below, this is always the case. While there could be other models for which higher-order terms in $m$ need to be considered, these are beyond the scope of this Letter. Moreover, for a transition to occur as $\lambda$ is varied, we need $\bar\sigma' = 0$, and $\lambda_\cc$ defined in Eq.~\eqref{eq:lc_def} has to be real-valued. This is the case if $\sigma(\rho)$ has a local minimum at $\rho = \bar\rho$, so that $\bar\sigma'' > 0$; otherwise, if $\bar\sigma'' < 0$, the bulk field has to be sufficiently strong so that
\begin{align} \label{eq:e_cond}
E^2 > \frac{2\pi^2\bar D^2}{\bar\sigma |\bar\sigma''|} \,.
\end{align}
We observe different transition behaviors depending on the sign of $\bar\sigma''$, each of which we discuss in the following.

\begin{figure*}
 \centering
  \includegraphics[width=0.95\textwidth]{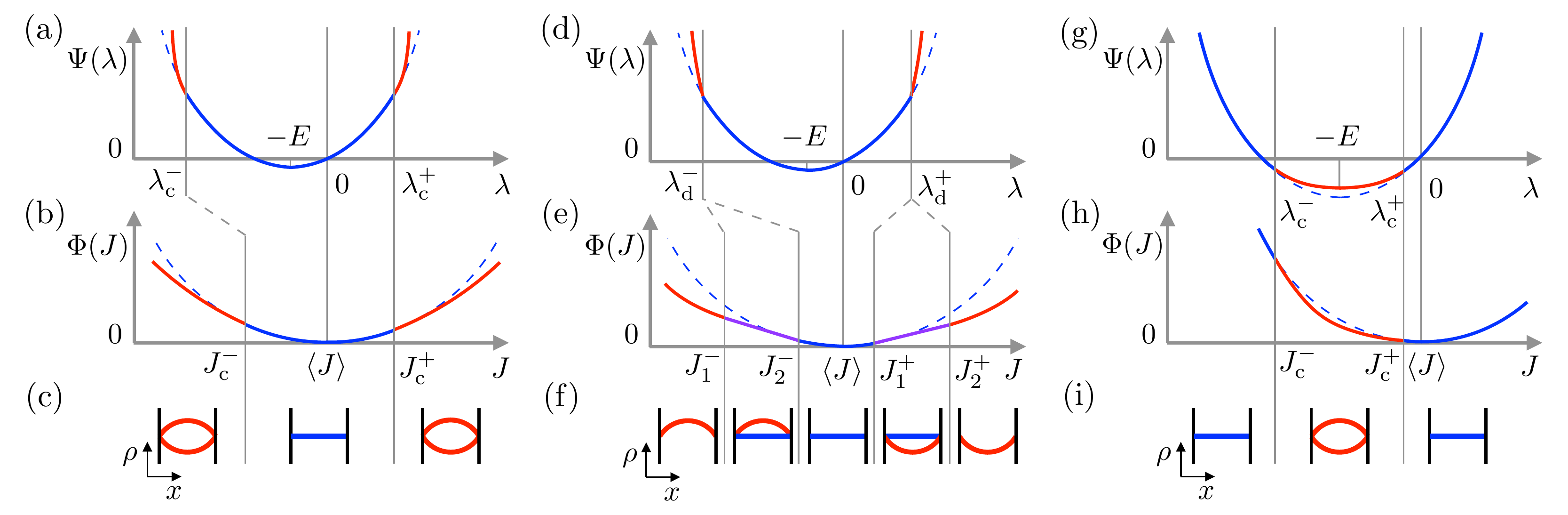}
  \caption{Schematic illustrations of the singularities and the optimal profiles for different types of phase transitions. The dashed (blue) lines represent behaviors of the functions if the Gaussian fluctuations persist for any $\lambda$ and $J$. Case 1: (a) The scaled CGF showing second-order singularities and (b) the corresponding LDF; (c) the shapes of optimal profiles as $J$ is varied; (d) The scaled CGF showing first-order singularities and (e) the corresponding LDF; (f) the optimal profiles as $J$ is varied. Case 2: (g) The scaled CGF showing second-order singularities, (h) the corresponding LDF, and (i) the optimal profiles as $J$ is varied.}
  \label{fig:fig1}
\end{figure*}

{\em Case 1a: $\bar\sigma'' > 0$, symmetry breaking} --- Consider a particle--hole symmetric system, whose Hamiltonian density shown in Eq.~\eqref{eq:hamiltonian} is invariant under the transformation defined by $x \to 1-x$, $\rho(x,t)  \to 2\bar\rho - \rho(1-x,t)$, and $\hat\rho(x,t) \to \lambda - \hat\rho(1-x,t)$. Assuming that $D(\rho)$ and $\sigma(\rho)$ are analytic, their odd-order derivatives vanish at $\rho = \bar\rho$, i.e.~$\bar D^{(2n+1)} = \bar\sigma^{(2n+1)} = 0$ for $n = 0,\,1,\,\ldots$ Then only the $m^2$ and $m^4$ terms survive in Eq.~\eqref{eq:landau_sym}, turning $\mathcal{L}(m)$ into the form of a Landau free energy of Ising-like systems. For $\lambda_\cc^- < \lambda < \lambda_\cc^+$, $\mathcal{L}(m)$ is minimized at $m^* = 0$, and $\Psi(\lambda)$ has a quadratic form corresponding to Gaussian fluctuations. For $\lambda > \lambda_\cc^+$ or $\lambda < \lambda_\cc^-$, we have $m^* \sim \pm |\lambda-\lambda_\cc|^{1/2}$, corresponding to a pair of symmetry-breaking profiles given by Eq.~\eqref{eq:defrhostar} which are mutually related by a particle--hole transformation defined above. This implies that for each instance of a current fluctuation $J$ in this regime, there is a symmetry breaking so that one of the two optimal profiles is observed with equal probability (see Fig.~\ref{fig:fig1}). Near the transition points, the scaled CGF $\Psi(\lambda)$ has singularities which behave as $\lim_{\lambda\downarrow\lambda_\cc}\Psi(\lambda) - \lim_{\lambda\uparrow\lambda_\cc}\Psi(\lambda)  \sim |\lambda - \lambda_\cc|^2$, implying second-order transitions. Clearly, the same critical scaling behavior is observed if $\bar D^{(3)}$, $\bar\sigma^{(5)}$ or higher-order derivatives are nonzero, although in such cases only one of the two density profiles is optimal.

{\em Case 1b: $\bar\sigma'' > 0$, first-order transition} --- Now consider the case when $\bar D'$ and $\bar\sigma^{(3)}$ have nonzero values. For a consistent Landau theory, we assume that $\bar D'$ and $\bar\sigma^{(3)}$ scale as $m^*$. Then the $m^3$ term induces a weak first-order singularity of the scaled CGF~\footnote{An exception could occur if the system is fine-tuned to satisfy $\bar D\bar\sigma^{(3)} = 3\bar D'\bar\sigma''$, see Appendix~\ref{ssec:landau_eq_asym}.}. On general grounds, similar results will be obtained even if $\bar D'$ and $\bar\sigma^{(3)}$ are larger. The transition shows up as jumps of $m^*$ at transition points $\lambda_\dd^\pm$ which are slightly shifted from $\lambda_\cc^\pm$, respectively (see Appendix~\ref{ssec:landau_eq_asym}). In a manner similar to Case 1a, the fluctuations are Gaussian for $\lambda_\dd^- < \lambda < \lambda_\dd^+$ and non-Gaussian otherwise (see Fig.~\ref{fig:fig1}). This behavior corresponds to a scenario discussed in \cite{bertini_non_2006}: when a current fluctuation $J$ occurs within the intervals $[J_1^\pm,J_2^\pm]$ defined by $J_1^\pm \equiv \lim_{\lambda \uparrow \lambda_\dd^\pm}\Psi'(\lambda)$ and $J_2^\pm \equiv \lim_{\lambda \downarrow \lambda_\dd^\pm}\Psi'(\lambda)$, we observe $J_1^\pm$ and $J_2^\pm$ with probability $p_1^\pm$ and $1-p_1^\pm$, respectively, such that $J = p_1^\pm J_1^\pm + (1-p_1^\pm) J_2^\pm$. This is a direct analog of phase coexistence in equilibrium first-order transitions.

{\em Case 2: $\bar\sigma'' < 0$} --- For Case 1, the bulk field $E$ is not essential for the existence of a DPT: it only shifts the location of the transition point according to Eq.~\eqref{eq:lc_def}. In contrast, for Case 2 phase transitions occur only when the bulk field $E$ is strong enough to satisfy Eq.~\eqref{eq:e_cond}. Since the form of $\mathcal{L}(m)$ remains the same, the system again exhibits symmetry breaking transitions for fully particle--hole symmetric systems, and first-order transitions in the absence of symmetry due to nonzero $\bar D'$ and $\bar\sigma^{(3)}$. Note that while the regions of non-Gaussian fluctuations were unbounded in Case 1, here they are bounded. This is because for $\bar\sigma'' < 0$ both transition points $\lambda_\cc^\pm$ have the same sign, as implied by Eq.~\eqref{eq:lc_def} (see Fig.~\ref{fig:fig1}).

{\em Generalization to $\rho_a \neq \rho_b$} --- We now turn to the case of unequal boundary densities given by $\bar\rho_a = \bar\rho - \delta\rho$ and $\bar\rho_b = \bar\rho + \delta\rho$. Treating $\delta\rho$ as a perturbation, we find to linear order in $\delta\rho$ that (see Appendix~\ref{sec:ueq_bcs})
\begin{align} \label{eq:psi}
\Psi(\lambda) = \frac{\bar\sigma}{2}\,\lambda\,(\lambda+2E) - 2\,\delta\rho \, \bar D \,\lambda - \inf_m \mathcal{L}(m) \,,
\end{align}
with only the quadratic term in $\mathcal{L}(m)$ modified as
\begin{align}
(\lambda - \lambda_\cc)m^2 \to \left(\lambda - \lambda_\cc - \frac{2\bar D}{\bar\sigma}\delta\rho\right)m^2,
\end{align}
which implies that the transition point is shifted but the other properties of phase transitions are unchanged. If $\bar\rho_a - \bar\rho \neq \bar\rho - \bar\rho_b$, we can use $(\bar\rho_a + \bar\rho_b)/2$ as the new value of $\bar\rho$. Provided that the odd-order derivatives of $D(\rho)$ and $\sigma(\rho)$ evaluated at the new $\bar\rho$ remain small, all results presented above are still valid.

{\em Microscopic models} --- We now present two lattice gas models, each of which exhibits one of the two cases of phase transitions described above.

{\em Case 1: $\bar\sigma'' > 0$} --- We consider a KLS~\cite{katz_nonequilibrium_1983} model with zero bulk bias, which features on-site exclusion and nearest-neighbor interactions. It is defined on a one-dimensional lattice, each site of which can be either occupied (``1'') or empty (``0''). The model is characterized by two parameters $\delta$ and $\varepsilon$, which govern the hopping dynamics according to the following transition rates (in arbitrary units):
\begin{align*}
&  0100\xrightarrow{1+\delta}0010 \,,\ 1101\xrightarrow{1-\delta
}1011\, ,\\
&  1100\xrightarrow{1+\varepsilon}1010 \,,\ 1010\xrightarrow{1-\varepsilon
}0110\, .
\end{align*}
Spatially inverted versions of these transitions occur with identical rates. Using the methods of \cite{Spohn_Book,hager_minimal_2001,krapivsky_unpub}, $D(\rho)$ and $\sigma(\rho)$ of the model can be derived exactly as functions of $\rho \in [0,1]$ (see Appendix~\ref{sec:kls} for their explicit forms). If $\delta = 0$, the model possesses a particle--hole symmetry, so that all odd-order derivatives of $D(\rho)$ and $\sigma(\rho)$ with respect to $\rho$ vanish at $\rho = 1/2$. More interestingly, for $\epsilon > 4/5$, one finds that $\sigma(1/2)$ becomes a local minimum. Thus, all results of Case 1a can be applied to this model by setting $\bar\rho = 1/2$. On the other hand, if $\delta \neq 0$, the system does not have a particle--hole symmetry. Then, for $\epsilon$ greater than some $\delta$-dependent threshold, $\sigma(\rho)$ has a local minimum at some $\delta$-dependent $\bar\rho$. All results of Case 1b are then applicable to this system.

{\em Case 2: $\bar\sigma'' < 0$} --- Consider a WASEP on a one-dimensional lattice of $L$ sites, whose hopping rates (in arbitrary units) are given by $10\xrightarrow{1+\delta}01$, $01\xrightarrow{1-\delta}10$. If $\delta = E/L$, it is well known~\cite{WASEP,gartner_convergence_1987} that the system is characterized by $D(\rho) = 1$ and $\sigma(\rho) = 2\rho(1-\rho)$, so that $\sigma''(\rho) < 0$ for any $\rho \in [0,1]$ with the maximum of $\sigma(\rho)$ located at $\rho = 1/2$. Applying the results of Case 2, Eq.~\eqref{eq:e_cond} implies that the system exhibits singularities of LDFs when $|E| > \pi$.

{\it Mechanism for symmetry breaking} --- To gain more intuition into the origin of the DPT, it is helpful to examine the Lagrangian formulation of the LDF~\cite{bertini_macroscopic_2015}
\begin{align} \label{eq:phi_lagrangian}
	\Phi(J) = \inf_{\rho} \int_0^1 \mathrm{d}x \, \frac{\left[J+D(\rho)\partial_x \rho - \sigma(\rho)E \right]^2}{2\sigma(\rho)} \,.
\end{align}
Close to the transition point, $\Phi(J)$ is minimized by an optimal profile of the form $\rho(x) = \bar\rho + m \sin(\pi x)$. Keeping the leading-order corrections in $m$, we obtain
\begin{align}
\Phi(J) &\simeq \frac{\delta J^2}{2\bar\sigma} \\
& + \inf_m \left[\left(\frac{\bar D^2}{2} - \frac{\bar\sigma''E \, \delta J}{2}	- \frac{\bar\sigma'' \, \delta J^2}{4\bar\sigma}\right)m^2 +O (m^4)\right] \,, \nonumber
\end{align}
where $\delta J \equiv J - \bar\sigma E$. The occurrence of symmetry breaking is controlled by the sign of the coefficient in front of~$m^2$, whose three terms represent contributions from diffusion, bulk field $E$, and noise amplitude. The first two originate from the numerator of Eq.~\eqref{eq:phi_lagrangian} and the last one comes from the denominator. The competition between these factors dictate whether it is beneficial to break the symmetry by density modulations. Depending on the sign of $\bar\sigma''$, there are two possible scenarios.

If $\bar\sigma'' > 0$, the coefficient of $m^2$ is positive for $\delta J$ close to zero and becomes negative for sufficiently large $\delta J$, signaling the symmetry breaking transition --- for large enough $\delta J$, the gain in action from the denominator overwhelms the cost of density modulations in the numerator. 

On the other hand, if $\bar\sigma'' < 0$, both the diffusion and the noise lead to a positive cost for density modulations. Negative contributions arise only from the field term. A large enough $E$ can make density modulations favorable for an intermediate range of $\delta J$, inducing a transition.

The origins of DPTs in these two cases are different. For $\bar\sigma'' > 0$ the transitions are due to the competition between the diffusion, which favors a flat profile, and the noise, which favors modulations. In contrast, for $\bar\sigma'' < 0$ the transitions are ruled by the contribution of the bulk field, which favors modulations, competing against the diffusion and the noise, both of which favor a flat profile. Similar arguments also apply to first-order transitions.

Comparisons with previous studies are in order. A recent study~\cite{Shpielberg2016} proposed a criterion which forbids DPTs of Case 2; but our results explicitly show that the WASEP is a counterexample to this criterion~\footnote{The argument for the criterion proposed by \cite{Shpielberg2016} does not apply to the case when $|\partial_x \hat\rho + E| < |E|$, which is true near the DPTs of Case 2. See Appendix~\ref{sec:stability_field} for more details.}. We also note that the asymmetric simple exclusion process (ASEP), which is non-diffusive, also exhibits DPTs in current fluctuations~\cite{dudzinski_relaxation_2000,deGier2005,gorissen_exact_2012,Lazarescu2015}. While these DPTs are remnants of the well-known boundary-induced transitions in mean behaviors, the DPTs of diffusive systems discussed above are very different.

There remains the question of how the DPTs discussed so far can be experimentally observed. Recently, the LDF for heat current in an RC circuit was empirically measured in~\cite{Ciliberto2013a,*Ciliberto2013b}, where the fast electronic dynamics allows the current LDF to be measured over a wide range~\footnote{Unlike the continuous media studied here, the systems studied in \cite{Ciliberto2013a,*Ciliberto2013b} are described by only a few degrees of freedom. Even for such finite-dimensional systems, the DPTs similar to those described in this Letter can still occur, as is to be discussed in \cite{baek_unpub}.}. To observe the DPTs discussed here in a similar experiment, one has to look at diffusive electronic transport with an extremum in $\sigma(\rho)$. These are common, resulting from non-monotonic changes in the electronic density of states. For example, minima of $\sigma(\rho)$ were observed in graphene transport~\cite{Adam2007,*Tan2007,*Chen2008} and maxima in fullerene peapods~\cite{Utko2010}. Using these systems, both cases of DPTs discussed above can in principle be observed.

In summary, we have studied a general one-dimensional diffusive transport through a channel connecting two reservoirs. Using a perturbative approach for general $D(\rho)$ and $\sigma(\rho)$, we find a large class of new DPTs which are not associated with the breaking of the additivity principle in the sense that the optimal profiles remain time-independent. For some of these DPTs we can explicitly prove the validity of the additivity principle, which we expect to hold for all cases (see Appendix~\ref{ssec:additivity}). It would be interesting to check whether other kinds of DPTs occur at larger values of $J$ or $\delta\rho$, and how the results can be generalized to higher dimensions.

\begin{acknowledgments}
We are grateful to Paul Krapivsky for his collaboration at early stages of this work. We also thank Giovanni Jona-Lasinio, Kirone Mallick, Ohad Shpielberg, Daniel Podolsky, and Michael Reznikov for helpful comments. YB and YK are supported by an ISF grant, and VL is supported by the ANR-15-CE40-0020-03 Grant LSD.	
\end{acknowledgments}

\onecolumngrid
\appendix

\section{Derivation of the saddle-point equations} \label{sec:saddle_point_eqns}

For completeness we outline the derivation of the saddle-point equations which are used to obtain the scaled cumulant generating function (CGF) for the time-averaged current. Similar derivations can also be found elsewhere in the literature (see, for example,~\cite{Imparato2009}).

From Eqs.~\eqref{eq:J_def} and \eqref{eq:scgf_def} of the main text, the scaled CGF is given by
\begin{align} \label{eq:cgf_path_avg}
\Psi(\lambda) = \lim_{T \to \infty} \frac{1}{T L} \ln \left\langle e^{\lambda L \int_0^T \mathrm{d}t\, \int_0^1\mathrm{d}x\, j(x,t)} \right\rangle \,,
\end{align}
where $\langle\cdot\rangle$ denotes an average over the noise realizations. Using the Langevin equation
\begin{align} \label{eq:langevin}
\partial_t \rho(x,t) + \partial_x j(x,t) = 0 \,, \quad
j(x,t)=-D(\rho)\partial_{x}\rho+\sigma(\rho)E+\sqrt{\sigma(\rho)}\eta(x,t)
\end{align}
with the spatial boundary conditions 
\begin{align} \label{eq:r_bcs}
\rho(0) = \bar\rho_a \,, \quad \rho(1) = \bar\rho_b \,,
\end{align}
the average on the r.h.s.~of Eq.~\eqref{eq:cgf_path_avg} can be written as
\begin{align} \label{eq:integ_rje}
\left\langle e^{\lambda L \int_0^{T}\mathrm{d}t\, \int_0^1\mathrm{d}x\, j} \right\rangle
= \left\langle \int \mathcal{D}\rho\, \mathcal{D}j\,
	 e^{\lambda L \int_0^{T}\mathrm{d}t\, \int_0^1\mathrm{d}x\, j}\, 
	\delta\left[\dot{\rho}+\nabla j\right] \,
	\delta\left[j + D(\rho)\nabla \rho - \sigma(\rho) E - \sqrt{\sigma(\rho)} \eta \right] \right\rangle\,.
\end{align}
The two delta functionals in the path integral make sure that the integration is carried out only over the paths governed by Eq.~\eqref{eq:langevin}. The functional $\delta\left[\dot{\rho}+\nabla j\right]$ can be rewritten in terms of its Fourier representation
\begin{align}
\delta\left[\dot{\rho}+\nabla j\right]
= \int \mathcal{D}\hat\rho \, e^{-L \int_0^T \mathrm{d}t \, \int_0^1 \mathrm{d}x \, \hat\rho(\dot{\rho}+\nabla j)} \,,
\end{align}
which introduces an auxiliary field $\hat{\rho}(x,t)$. Then Eq.~\eqref{eq:integ_rje} can be integrated over the current $j(x,t)$ and the noise $\eta(x,t)$ to yield
\begin{align}
\left\langle e^{\lambda L \int_0^{T}\mathrm{d}t\, \int_0^1\mathrm{d}x\, j} \right\rangle
&= \int \mathcal{D}\rho\, \mathcal{D}\hat\rho
	\exp \left\{L \int_0^{T}\mathrm{d}t\, 
	\int_0^1\mathrm{d}x\,
	\left[-\hat\rho \dot{\rho} - D(\rho)(\nabla \rho) (\lambda + \nabla\hat\rho)
	+\frac{\sigma(\rho)}{2}(\lambda + \nabla \hat\rho)(\lambda + \nabla\hat\rho + 2E) \right] \right\} \,.
\end{align}
Here the auxiliary field variable $\hat\rho(x,t)$ satisfies the boundary conditions
\begin{align} \label{eq:rh_bcs}
\hat\rho(0,t) = 0,\quad \hat\rho(1,t) = 0,
\end{align}
which accounts for the absence of fluctuations at the boundaries~\cite{Tailleur2008}.  It is useful to introduce a change of variables
\begin{align} \label{eq:rh2rhl}
\hat\rho(x,t) \to \hat\rho_\lambda(x,t) - \lambda x,
\end{align}
which gives
\begin{align} \label{eq:integ_rrh}
\left\langle e^{\lambda L \int_0^{T}\mathrm{d}t\, \int_0^1\mathrm{d}x\, j} \right\rangle
&= \int \mathcal{D}\rho\, \mathcal{D}\hat\rho_\lambda
	\exp \left\{- L \int_0^{T}\mathrm{d}t\, 
	\int_0^1\mathrm{d}x\,
	\left[\hat\rho_\lambda \dot{\rho} + D(\rho)(\nabla \rho) (\nabla\hat\rho_\lambda)
	-\frac{\sigma(\rho)}{2}(\nabla \hat\rho_\lambda)(\nabla\hat\rho_\lambda + 2E) \right] \right\}.
\end{align}
Here a temporal boundary term $L\int_0^1\mathrm{d}x\, \{\lambda x [\rho(x,T)-\rho(x,0)]\}$ in the exponent is neglected as it becomes negligible for $T \gg 1$.

Since we are interested in the large $L$ limit, the scaled CGF can be evaluated using a saddle point so that
\begin{align} \label{eq:psi_least_action}
\Psi(\lambda) = - \lim_{T \to \infty} \frac{1}{T} \inf_{\rho,\,\hat\rho_\lambda} \int_0^{T}\mathrm{d}t\, 
	\int_0^1\mathrm{d}x\, \left[ \hat\rho_\lambda \dot{\rho}  - H(\rho,\hat\rho_\lambda)\right],
\end{align}
with the Hamiltonian density $H$ defined as
\begin{align}
H(\rho,\hat\rho_\lambda) \equiv -D(\rho)(\nabla \rho) (\nabla\hat\rho_\lambda)
	+\frac{\sigma(\rho)}{2}(\nabla \hat\rho_\lambda)(\nabla\hat\rho_\lambda + 2E) \,.
\end{align}
Here $\rho$ and $\hat\rho_\lambda$ can be interpreted as position and momentum variables, respectively. Thus, the saddle-point solutions are obtained by solving the equations
\begin{align} \label{eq:saddle_traj}
\dot{\rho} = \nabla\left[D(\rho)\nabla\rho - \sigma(\rho)(\nabla\hat\rho_\lambda +E)\right], \quad
\dot{\hat\rho}_\lambda = - D(\rho) \nabla^2 \hat\rho_\lambda - \frac{1}{2}\sigma'(\rho)(\nabla\hat\rho_\lambda)(\nabla\hat\rho_\lambda + 2E)
\end{align}
with the spatial boundary conditions
\begin{align}
\rho(0,t) = \bar\rho_a,\quad \rho(1,t) = \bar\rho_b,\quad \hat\rho_\lambda(0,t) = 0,\quad \hat\rho_\lambda(1,t) = \lambda \;.
\end{align}
These are consistent with Eqs.~\eqref{eq:r_bcs}, \eqref{eq:rh_bcs}, and \eqref{eq:rh2rhl}. With the understanding that these boundary conditions are always assumed, for brevity in what follows we drop the subscript $\lambda$ from $\hat\rho_\lambda$.

\section{Equal boundary densities} \label{sec:eq_bcs}

In what follows, we first consider the case when the two particle reservoirs have equal densities $\bar\rho_a = \bar\rho_b = \bar\rho$. The case $E=0$ then corresponds to an equilibrium system, while when $E \neq 0$ the system is out of equilibrium. We identify the symmetry-breaking transition point $\lambda_\cc$ discussed in the main text, and then construct from first principles a Landau theory for the transition. Finally, we prove that in this case the additivity principle (assumed to hold throughout the derivation) is valid.

\subsection{Symmetry breaking in particle--hole symmetric systems}
\label{ssec:sym_break_eq}

We start by analyzing systems which are particle--hole symmetric about $\rho = \bar\rho$, so that all odd-order derivatives of the transport coefficients at $\rho = \bar\rho$ vanish:
\begin{align} \label{eq:odd_deriv_zero}
\bar D^{(2n+1)} \equiv D^{(2n+1)}(\bar\rho) = 0 \,, \quad
\bar\sigma^{(2n+1)} \equiv \sigma^{(2n+1)}(\bar\rho) = 0 \quad
\text{for $n = 0$, $1$, $2$, $\ldots$}
\end{align}
Assuming that the additivity principle holds, the saddle-point equations~\eqref{eq:saddle_traj} reduce to their time-independent forms
\begin{align} \label{eq:saddle_traj_time_indept}
\nabla\left[D(\rho)\nabla\rho - \sigma(\rho)(\nabla\hat\rho +E)\right] = 0 \,, \quad
D(\rho) \nabla^2 \hat\rho + \frac{1}{2}\sigma'(\rho)(\nabla\hat\rho)(\nabla\hat\rho+ 2E) = 0 \,.
\end{align}
As discussed in the main text, such system may exhibit a dynamical phase transition at $\lambda=\lambda_\cc$. For $\lambda<\lambda_\cc$ the density and momentum profiles
\begin{align} \label{eq:rho_sym_eq}
\rho^\text{sym}(x) \equiv \bar\rho \,, \quad
\hat\rho^\text{sym}(x) \equiv \lambda x \,,
\end{align}
which are symmetric in the sense that they are invariant under
\begin{align} \label{eq:par_hole_mapping}
\rho(x) \to 2\bar\rho - \rho(1-x) \,, \quad \hat\rho(x) \to \lambda - \hat\rho(1-x)\,,
\end{align}
are the only solution to the saddle-point equation. In contrast, for $\lambda>\lambda_\cc$ two additional symmetry-breaking solutions appear and become more dominant than the symmetric profile. In what follows we prove the existence of this transition.

Note that near a transition point $\lambda = \lambda_\cc$, the symmetry of the optimal profile is weakly broken by small deviations from Eq.~\eqref{eq:rho_sym_eq}. In other words,
\begin{align} \label{eq:another_saddle}
\rho(x) = \rho^\text{sym}(x) + \varphi(x) \,, \quad
\hat\rho(x) = \hat\rho^\text{sym}(x) + \hat\varphi(x)
\end{align}
with small but nonzero $\varphi$ and $\hat\varphi$ satisfying the boundary conditions
\begin{align} \label{eq:phi_zero_bcs}
\varphi(0) = \varphi(1) = \hat\varphi(0) = \hat\varphi(1) = 0
\end{align}
will be another solution of Eq.~\eqref{eq:saddle_traj_time_indept}. Linearizing Eq.~\eqref{eq:saddle_traj_time_indept} with respect to $\varphi$ and $\hat\varphi$, we obtain a system of linear differential equations
\begin{align} \label{eq:phi_saddle_time_indept}
\bar D \nabla^2 \varphi - \bar\sigma \nabla^2 \hat\varphi = 0, \quad
\bar D \nabla^2 \hat\varphi + \frac{\bar\sigma''}{2}\lambda(\lambda+2E) \varphi = 0.
\end{align}
Using the Fourier transforms
\begin{align} \label{eq:fourier_time_indept}
\varphi(x) = \sum_{n = 1}^\infty \psi_n \sin(n\pi x), \quad
\hat\varphi(x) = \sum_{n = 1}^\infty \hat\psi_n \sin(n\pi x),
\end{align}
we can rewrite Eq.~\eqref{eq:phi_saddle_time_indept} as
\begin{align} \label{eq:psi_n}
\bar D \psi_n - \bar\sigma \hat\psi_n = 0,
\quad
n^2 \pi^2 \bar D \hat\psi_n - \frac{\bar\sigma''}{2}\lambda(\lambda+2E)\psi_n = 0.
\end{align}
These linear equations have nonzero solutions for $\psi_n$ and $\hat\psi_n$ if and only if
\begin{align}
\lambda(\lambda+2E) = \frac{2 n^2 \pi^2 \bar D^2}{\bar\sigma \bar\sigma''} \,,
\end{align}
i.e.~when $\lambda$ is equal to
\begin{align} \label{eq:lambda_cn}
\lambda_{\text{c},n}^\pm \equiv -E \pm \sqrt{E^2 + \frac{2n^2 \pi^2 \bar D^2}{\bar\sigma \bar\sigma''}} \,.
\end{align}
From Eq.~\eqref{eq:psi_n} it is clear that if $\psi_n$ and $\hat\psi_n$ are solutions, so are $-\psi_n$ and $-\hat\psi_n$. This implies that Eq.~\eqref{eq:saddle_traj_time_indept} allows a symmetry breaking at critical values $\lambda_{\text{c},n}^\pm$. The symmetry breaking transition will clearly occur for the $n$ with a minimal $|\lambda_{\text{c},n}^\pm|$. For the case $\bar\sigma'' > 0$, we always have $\lambda_{\text{c},n}^- < 0 < \lambda_{\text{c},n}^+$, so that the symmetry breaking occurs on both sides of $\lambda = 0$. It is clear then that $|\lambda_{\text{c},n}^\pm|$ is minimized for $n = 1$.

For the case $\bar\sigma'' < 0$, we assume that $|E|$ is large enough to keep $\lambda_{\text{c},n}^\pm$ real-valued, as discussed in the main text. If $E > 0$ ($E < 0$), we have $\lambda_{\text{c},n}^- < \lambda_{\text{c},n}^+ < 0$ ($0 < \lambda_{\text{c},n}^- < \lambda_{\text{c},n}^+$), which means that the symmetry breaking occurs as $\lambda$ is decreased from zero to $\lambda_{\text{c},n}^+$ (increased from zero to $\lambda_{\text{c},n}^-$). Again, $|\lambda_{\text{c},n}^+|$ ($|\lambda_{\text{c},n}^-|$) is minimized at $n = 1$.

Hence, regardless of the sign of $\bar\sigma''$, only a deviation of the form $\varphi(x) \sim \sin(\pi x)$ is relevant to the symmetry-breaking transition. Moreover, due to the Gallavotti--Cohen symmetry~\cite{Gallavotti1995a,*Gallavotti1995b}
\begin{align}
\Psi(\lambda) = \Psi(- 2E - \lambda),
\end{align}
if a dynamical phase transition occurs at $\lambda = \lambda_{\text{c},1}^+$, the same kind of transition occurs at $\lambda = \lambda_{\text{c},1}^-$, and vice versa. Thus the transition points are always at
\begin{align} \label{eq:lambda_c_eq}
\lambda_\cc^\pm \equiv \lambda_{\text{c},1}^\pm = -E \pm \sqrt{E^2 + \frac{2 \pi^2 \bar D^2}{\bar\sigma \bar\sigma''}} \,.
\end{align}
In the following, for simplicity we will use $\lambda_\cc$ when we are referring to one of the two transition points.

While we have shown that another solution appears at $\lambda_\cc$ we have not shown that it dominates to CGF thus leading to a transition. This is done next by deriving a Landau theory for the transition from first principles.

\subsection{Landau theory for symmetry-breaking transitions} \label{ssec:landau_eq_sym}

Here we give a detailed derivation of the Landau theory describing the symmetry-breaking transitions. For this purpose, the system is again assumed to be particle--hole symmetric, so that Eqs.~\eqref{eq:odd_deriv_zero} and \eqref{eq:rho_sym_eq} of Sec.~\ref{ssec:sym_break_eq} are still valid.

Under the assumption of the additivity principle, the scaled CGF can be obtained from a time-independent version of Eq.~\eqref{eq:psi_least_action}:
\begin{align} \label{eq:psi_least_action_time_indept}
\Psi(\lambda) = \sup_{\rho,\,\hat\rho} \int_0^1 \mathrm{d}x\, H(\rho,\hat\rho).
\end{align}
To construct a Landau theory of the transition, for $\lambda$ close to $\lambda_\cc$ we can use an expansion 
\begin{align} \label{eq:phi_opt}
\varphi(x) &= m^* \sin (\pi x) + (m^*)^2 \varphi_2(x)
+ (m^*)^3 \varphi_3(x) + O\left[(m^*)^4\right]\,, \nonumber\\
\hat\varphi(x) &= m^* \frac{\bar D}{\bar\sigma}\sin (\pi x) + (m^*)^2 \hat\varphi_2(x)
+ (m^*)^3 \hat\varphi_3(x) + O\left[(m^*)^4\right] \,,
\end{align}
where $m^*$ measures the contribution of $\sin(\pi x)$ to the symmetry breaking. The functions  $\varphi_2$, $\varphi_3$, $\ldots$ and $\hat\varphi_2$, $\hat\varphi_3$, $\ldots$ are zero at the boundaries ($x = 0$ and $x = 1$) and orthogonal to $\sin (\pi x)$; they are to be determined by solving Eq.~\eqref{eq:saddle_traj_time_indept} perturbatively (see below). Then, if we define 
\begin{align} \label{eq:phi_m}
\varphi^m(x) &\equiv m \sin (\pi x) + m^2 \varphi_2(x)
+ m^3 \varphi_3(x) + O(m^4) \,, \nonumber\\
\hat\varphi^m(x) &\equiv m \frac{\bar D}{\bar\sigma}\sin (\pi x) + m^2 \hat\varphi_2(x)
+ m^3 \hat\varphi_3(x) + O(m^4) \,,
\end{align}
a Landau function can be written as
\begin{align} \label{eq:landau_def}
{\cal L}(m) \equiv \int_0^1 \mathrm{d}x\, \left[
	H(\rho^\text{sym},\hat\rho^\text{sym})
	- H(\rho^\text{sym} + \varphi^m,\hat\rho^\text{sym} + \hat\varphi^m) \right] \,,
\end{align}
and the minimization problem associated with the scaled CGF can be cast as 
\begin{align}
\Psi(\lambda) = \int_0^1 \mathrm{d}x\, H(\rho^\text{sym},\hat\rho^\text{sym})
	- \inf_m {\cal L}(m) \,.
\end{align}
Thus here a minimization over profiles in Eq.~\eqref{eq:psi_least_action_time_indept} is simplified to that over a single parameter $m$.

To calculate $\varphi_2$, $\varphi_3$, $\ldots$ and $\hat\varphi_2$, $\hat\varphi_3$, $\ldots$, we substitute $\rho = \rho^\text{sym} + \varphi$ and $\hat\rho = \hat\rho^\text{sym} + \hat\varphi$ into Eq.~\eqref{eq:saddle_traj_time_indept} and expand the equations with respect to $m^*$. This allows us to solve the differential equations order by order. The perturbation analysis can be carried out in a well-defined way if the distance from the symmetry-breaking transition point $\delta\lambda \equiv \lambda - \lambda_\cc$ satisfies a scaling relation with $m^*$. Inspired by an Ising Landau theory, we use the scaling ansatz
\begin{align} \label{eq:scalings}
\delta\lambda \simeq c^{\delta\lambda} (m^*)^2,
\end{align}
where the value of the coefficient $c^{\delta\lambda}$ is determined below. Carrying out this procedure to order $(m^*)^2$, we obtain
\begin{align} \label{eq:ode_20}
\nabla^2 \varphi_2 = -\pi^2 \varphi_2, \quad
\nabla^2 \hat\varphi_2 = \frac{\bar D}{\bar\sigma} \nabla^2 \varphi_2 - \frac{\pi\bar\sigma''(\lambda_\cc+E)}{2\bar\sigma}\sin(2\pi x).
\end{align}
Keeping in mind that $\varphi_2$ must be orthogonal to $\sin(\pi x)$ gives
\begin{align}
\varphi_2(x) = 0, \quad
\hat\varphi_2(x) = \frac{\bar\sigma''(\lambda_\cc+E)}{8\pi \bar\sigma}\sin(2\pi x).
\end{align}
To order $(m^*)^3$, we obtain
\begin{align} \label{eq:ode_30_1}
\nabla^2 \varphi_3 &= -\pi^2 \varphi_3 + \left[\frac{1}{8}\left(\frac{4\pi^2 \bar D''}{\bar D} + \frac{\bar\sigma''^2 E^2}{\bar D^2} - \frac{\pi^2 \bar\sigma^{(4)}}{\bar\sigma''}\right) - \frac{(\lambda_\cc+E) \bar\sigma\bar\sigma'' c^{\delta\lambda}}{\bar D^2}\right] \sin (\pi  x) \nonumber\\
	&\quad -\frac{1}{24} \left(\frac{12 \pi ^2 \bar D''}{\bar D}+\frac{3 \bar\sigma''^2 E^2}{\bar D^2}-\frac{\pi ^2 \bar\sigma^{(4)}}{\bar\sigma''}\right) \sin (3 \pi  x)
\end{align}
and
\begin{align} \label{eq:ode_30_2}
\nabla^2 \hat\varphi_3 &= \frac{\bar D}{\bar\sigma} \nabla^2 \varphi_3
	+\frac{\pi ^2 \left(\bar D \bar\sigma''-\bar\sigma \bar D''\right)}{8 \bar\sigma^2} \left[\sin (\pi  x) - 3\sin (3\pi x)\right].
\end{align}
The differential equation \eqref{eq:ode_30_1} has terms of the form
\begin{align} \label{eq:ode_f}
\nabla^2 f(x) = -\pi^2 f(x) + a \sin(\pi x) \,.
\end{align}
This equation has a solution with $f(0) = f(1) = 0$ if and only if $a = 0$. This condition fixes the coefficient
\begin{align} \label{eq:ode_30_const}
c^{\delta\lambda} = \frac{\bar D^2}{8 (\lambda_\cc + E) \bar\sigma \bar\sigma''^2}\left(\frac{4\pi^2 \bar D''}{\bar D} + \frac{\bar\sigma''^2 E^2}{\bar D^2} - \frac{\pi^2 \bar\sigma^{(4)}}{\bar\sigma''}\right),
\end{align}
with which we obtain the solutions
\begin{align}
\varphi_3(x) &= \frac{1}{192\pi^2}\left(\frac{12\pi^2\bar D''}{\bar D}+\frac{3\bar\sigma''^2 E^2}{\bar D^2}-\frac{\pi^2\bar\sigma^{(4)}}{\bar\sigma''}\right)\sin (3 \pi  x)
\end{align}
and
\begin{align}
\hat\varphi_3(x) &= - \frac{\bar D \bar\sigma''-\bar D''\bar\sigma}{8\bar\sigma^2}\sin (\pi  x)
	+\left[\frac{\bar D}{192\pi^2\bar\sigma}\left(\frac{12\pi^2\bar D''}{\bar D}+\frac{3\bar\sigma''^2 E^2}{\bar D^2}-\frac{\pi^2\bar\sigma^{(4)}}{\bar\sigma''}\right) + \frac{\bar D \bar\sigma''-\bar D'' \bar\sigma}{24\bar\sigma^2}\right]\sin (3 \pi  x) \,.
\end{align}

Using $\varphi^m$ and $\hat\varphi^m$ to order $m^3$, we finally obtain
\begin{align} \label{eq:landau_eq_sym}
{\cal L}(m) &= 
	-\frac{(\lambda_\cc + E) \bar\sigma''}{4} \,\delta\lambda \, m^2
	+ \left[\frac{\pi^2\bar D\left(4\bar D'' \bar\sigma'' - \bar D \bar\sigma^{(4)}\right)}{64\bar\sigma\bar\sigma''} + \frac{\bar\sigma''^2 E^2}{64\bar\sigma}
	\right] m^4 + O(m^5) \,,
\end{align}
which indeed has the form of a Landau function describing a symmetry-breaking transition at $\delta\lambda = 0$. We note that Eqs.~\eqref{eq:scalings} and \eqref{eq:ode_30_const} guarantee ${\cal L}'(m^*) = 0$, which is indeed a condition required for the optimal value of the order parameter $m$.

\subsection{Systems without particle--hole symmetry} \label{ssec:landau_eq_asym}

The derivation of the Landau theory described above can be generalized to systems with a weak particle--hole asymmetry. To construct a consistent perturbative Landau function, we have to take the odd-order derivatives $\bar D'$, $\bar\sigma'$, and $\bar\sigma^{(3)}$ (which contribute to the asymmetry) to scale with $m^*$ (as was done for $\delta\lambda$ in the previous discussion). More specifically, we now assume
\begin{align} \label{eq:scalings_asym}
\delta\lambda \simeq c^{\delta\lambda} (m^*)^2 \,, \quad
\bar \sigma' \simeq c^{\bar\sigma'} (m^*)^3 \,, \quad
\bar D' \simeq c^{\bar D'} m^* \,, \quad
\bar \sigma^{(3)} \simeq c^{\bar\sigma^{(3)}} m^* \,.
\end{align}
Then we can again put $\rho = \rho^\text{sym} + \varphi$ and $\hat\rho = \hat\rho^\text{sym} + \hat\varphi$ into Eq.~\eqref{eq:saddle_traj_time_indept}, expand the equations, and solve them with the boundary conditions order by order. The equations can be solved only if 
\begin{align} \label{eq:ode_30_const_b}
c^{\delta\lambda} = \frac{4 \pi\bar D^2}{(\lambda_\cc + E) \bar\sigma \bar\sigma''}\left(\frac{c^{\bar D'}}{\bar D}-\frac{3 c^{\bar\sigma'} + c^{\bar\sigma^{(3)}}}{3\bar\sigma''}\right)+\frac{\bar D^2}{8 (\lambda_\cc + E) \bar\sigma \bar\sigma''^2}\left(\frac{4\pi^2 \bar D''}{\bar D} + \frac{\bar\sigma''^2 E^2}{\bar D^2} - \frac{\pi^2 \bar\sigma^{(4)}}{\bar\sigma''}\right).
\end{align}
This relation does not mean that only three parameters among $\delta\lambda$, $\bar D'$, $\bar\sigma'$, and $\bar\sigma^{(3)}$ are mutually independent. The degree of freedom which is actually lost is $m^*$, whose value is obtained by combining Eqs.~\eqref{eq:scalings} and \eqref{eq:ode_30_const_b}. 

Finally, using the solutions for $\varphi^m$ and $\hat\varphi^m$ up to the order of $m^3$, we can combine Eqs.~\eqref{eq:landau_def} and \eqref{eq:phi_m} to obtain
\begin{align} \label{eq:landau_eq_asym}
{\cal L}(m) &= 
	-\frac{2\pi \bar D^2}{\bar\sigma\bar\sigma''} \, \bar\sigma' \, m
	-\frac{(\lambda_\cc + E) \bar\sigma''}{4} \,\delta\lambda \, m^2
	-\frac{2\pi \bar D(\bar D \bar\sigma^{(3)}-3\bar D'\bar\sigma'')}{9\bar\sigma\bar\sigma''}
	\,m^3 \nonumber\\
	&\quad + \left[\frac{\pi^2\bar D\left(4\bar D'' \bar\sigma'' - \bar D \bar\sigma^{(4)}\right)}{64\bar\sigma\bar\sigma''} + \frac{\bar\sigma''^2 E^2}{64\bar\sigma}
	\right] m^4 + O(m^5),
\end{align}
which is the expression for the Landau function presented in Eq.~(14) of the main text. If $\bar\sigma' = 0$, $\bar D\bar\sigma^{(3)} \neq 3\bar D'\bar\sigma''$, and the coefficient of $m^4$ is positive, this Landau function implies discontinuous transitions at
\begin{align}
	\lambda_\dd^\pm = \lambda_\cc^\pm - \frac{128(\lambda_\cc^\pm + E)}{27\bar\sigma''}\frac{(3\bar D'\bar\sigma''-\bar D\bar\sigma^{(3)})^2}{\pi^2 \bar D(4\bar D''\bar\sigma'' - \bar D\bar \sigma^{(4)}) + \bar\sigma''^3 E^2}\,.
\end{align}
If $\bar D\bar\sigma^{(3)} = 3\bar D'\bar\sigma''$, the Landau function becomes identical to Eq.~\eqref{eq:landau_eq_sym} up to order $m^4$. This means that the system exhibits continuous transitions at $\lambda = \lambda_\cc^\pm$, which are no longer symmetry-breaking transitions because the particle--hole symmetry is already broken by higher-order terms. Finally, we note that Eqs.~\eqref{eq:scalings} and \eqref{eq:ode_30_const_b} guarantee ${\cal L}'(m^*) = 0$, which is indeed a condition required for $m^*$.

\subsection{Validity of the additivity principle} \label{ssec:additivity}

So far we assumed that the additivity principle holds. One might be worried about possible time-dependent saddle-point solutions with a lower action. Here we prove that this is not the case for systems with equal boundary densities and a particle--hole symmetry. This is done by studying time-dependent perturbations of the symmetric profile given by
\begin{align}
\rho(x,t) = \rho^\text{sym}(x) + \varphi(x,t) \,, \quad \hat\rho(x,t) = \hat\rho^\text{sym}(x) + \hat\varphi(x,t)\,,
\end{align}
with the boundary conditions
\begin{align}
\varphi(0,t) = \varphi(1,t) = \hat\varphi(0,t) = \hat\varphi(1,t) = 0 \,.
\end{align}

As the first step, we linearize the time-dependent saddle-point equations~\eqref{eq:saddle_traj} with respect to these perturbations, which gives
\begin{align} \label{eq:phi_saddle_traj_eq}
\dot{\varphi} = \bar D \nabla^2 \varphi - \bar\sigma \nabla^2 \hat\varphi \,, \quad
\dot{\hat\varphi} = - \bar D \nabla^2 \hat\varphi - \frac{\bar\sigma''}{2}\lambda(\lambda+2E) \varphi \,.
\end{align}
Using the Fourier transforms
\begin{align} \label{eq:phi_fourier}
\varphi(x,t) = \sum_{n = 1}^\infty \int_{-\infty}^\infty \frac{\mathrm{d}\omega}{2\pi} \,
	\psi_n(\omega) \, e^{i \omega t} \sin(n\pi x) \,, \quad
\hat\varphi(x,t) = \sum_{n = 1}^\infty \int_{-\infty}^\infty \frac{\mathrm{d}\omega}{2\pi} \,
	\hat{\psi}_n(\omega) \, e^{i \omega t} \sin(n\pi x) \,,
\end{align}
we can rewrite Eq.~\eqref{eq:phi_saddle_traj_eq} as
\begin{align}
i\omega\psi_n(\omega) = - n^2 \pi^2 \bar D \psi_n(\omega) + n^2 \pi^2 \bar\sigma \hat\psi_n(\omega) \,,
\quad
i\omega\hat\psi_n(\omega) = n^2 \pi^2 \bar D \hat\psi_n(\omega) - \frac{\bar\sigma''}{2}\lambda(\lambda+2E)\psi_n(\omega) \,.
\end{align}
These linear equations have nonzero solutions for $\psi_n$ and $\hat\psi_n$ if and only if
\begin{align} \label{eq:lambda_sym_break}
\lambda(\lambda+2E) = \frac{2 \left(n^4 \pi^4 \bar D^2 + \omega^2\right)}{n^2 \pi^2 \bar\sigma \bar\sigma''} \,.
\end{align}
This implies that Eq.~\eqref{eq:saddle_traj} allows a symmetry breaking by $\varphi(x,t) \sim e^{i\omega t}\sin(n\pi x)$ if $\lambda$ is equal to
\begin{align} \label{eq:lambda_cn_omega}
\lambda_{\text{c},n}^\pm(\omega) \equiv -E \pm \sqrt{E^2 + \frac{2 \left(n^4 \pi^4 \bar D^2 + \omega^2\right)}{n^2 \pi^2 \bar\sigma \bar\sigma''}} \,,
\end{align}
with depends on both $n$ and $\omega$.

The rest of the proof is a repetition of the argument by which we identified the symmetry-breaking transition point in Sec.~\ref{ssec:sym_break_eq}. The transition occurs for the values of $n$ and $\omega$ which minimize $|\lambda_{\text{c},n}^\pm(\omega)|$. Since increasing $|\omega|$ has the same effect on $|\lambda_{\text{c},n}^\pm(\omega)|$ as increasing $n$ does, both parameters have the smallest possible value at the transition point, so that $n = 1$ and $\omega = 0$. This implies that the symmetry-breaking profile has the longest possible wavelength ($n = 1$) and zero frequency ($\omega =0$). The result is consistent with the value of $\lambda_\cc$ obtained in Eq.~\eqref{eq:lambda_c_eq}. This shows that the additivity principle is valid at the transition point.

\section{Unequal boundary densities} \label{sec:ueq_bcs}

We now discuss the case when the two particle reservoirs have unequal densities $\bar\rho_a = \bar\rho - \delta\rho$ and $\bar\rho_b = \bar\rho + \delta\rho$, so that the system has a boundary driving in addition to the possible bulk driving. Assuming the boundary driving to be small ($\delta\rho \ll 1$), we perturbatively obtain the linear corrections to the results obtained in Sec.~\ref{sec:eq_bcs}.

\subsection{Symmetry-breaking transition point and the additivity principle}

We first discuss how the transition point $\lambda_\cc$ and the validity of the additivity principle are affected by the boundary driving $\delta\rho$. If $\delta\rho \neq 0$, the symmetric density and momentum profiles $\rho^\text{sym}(x)$ and $\hat\rho^\text{sym}(x)$ given by Eqs.~\eqref{eq:phi_opt} and~\eqref{eq:phi_m} are no longer valid saddle-point solutions, because they are inconsistent with the boundary conditions for $\bar\rho_a$ and $\bar\rho_b$. Thus there must be corrections which alter the symmetric profiles as
\begin{align} \label{eq:rho_sym_ueq}
\rho^\text{sym}(x) \equiv \bar\rho + \delta\rho\,\rho_1(x) + O(\delta\rho^2) \,, \quad
\hat\rho^\text{sym}(x) \equiv \lambda x + \delta\rho\,\hat\rho_1(x) + O(\delta\rho^2) \,.
\end{align}
Solving the saddle-point equations~\eqref{eq:saddle_traj_time_indept} perturbatively, the linear corrections are obtained as
\begin{align} \label{eq:rho_sym_ueq_corr}
\rho_1(x) = \csc \frac{\alpha(\lambda)}{2} \sin \left[\alpha(\lambda)\left(x-\frac{1}{2}\right)\right] \,, \quad 
\hat\rho_1(x) = \frac{\bar D}{\bar \sigma} \rho_1(x) - \frac{2\bar D}{\bar\sigma} \left( x - \frac{1}{2}\right) \,,
\end{align}
with $\alpha(\lambda)$ denoting
\begin{align} \label{eq:alpha}
\alpha(\lambda) \equiv \sqrt{\frac{\lambda(\lambda+2E)\bar\sigma\bar\sigma''}{2\bar D^2}} \,.
\end{align}
It is easy to verify that the profiles given by Eqs.~\eqref{eq:rho_sym_ueq} and \eqref{eq:rho_sym_ueq_corr} are indeed symmetric under Eq.~\eqref{eq:par_hole_mapping}.

Based on the modified symmetric profiles obtained above, we identify the critical $\lambda$ at which symmetry-breaking saddle-point solutions are allowed. This can be done by repeating the procedure described in Sec.~\ref{ssec:additivity} while keeping track of the linear corrections in $\delta\rho$. After some algebra, we find that Eq.~\eqref{eq:lambda_cn_omega} is modified to
\begin{align}
\lambda_{\text{c},n}^\pm(\omega) \simeq -E \pm \sqrt{E^2 + \frac{2 \left(n^4 \pi^4 \bar D^2 + \omega^2\right)}{n^2 \pi^2 \bar\sigma \bar\sigma''}} + \frac{2 \bar D}{\bar \sigma}\delta\rho \,,
\end{align}
which shows that up to order $\delta\rho$ the threshold is shifted by the same amount for each value of $n$ and $\omega$. As already discussed, the actual symmetry-breaking transition occurs for the values of $n$ and $\omega$ which minimize $|\lambda_{\text{c},n}^\pm(\omega)|$. Thus the transition occurs at the critical point given by the longest wavelength time-independent deviation ($n = 1$ and $\omega = 0$), as in the case of $\delta\rho = 0$. Thus the additivity principle remains valid up to order $\delta\rho$, and the transition point is shifted by
\begin{align} \label{eq:lambda_c_ueq}
\lambda_\cc \to \lambda_\cc + \frac{2 \bar D}{\bar \sigma}\delta\rho \,.
\end{align}

\subsection{Derivation of the Landau theory}

The Landau theory for $\delta\rho \neq 0$ can be derived through a procedure which is almost the same as the one for $\delta\rho = 0$ described in Sec.~\ref{ssec:landau_eq_sym} and \ref{ssec:landau_eq_asym}, except that we need to keep track of the linear corrections in $\delta\rho$. These corrections appear in the symmetric profiles $\rho^\text{sym}$ and $\hat\rho^\text{sym}$ as obtained in Eq.~\eqref{eq:rho_sym_ueq_corr}, the deviations of the optimal profiles $\varphi$ and $\hat\varphi$ introduced in Eq.~\eqref{eq:phi_opt}, and the amplitudes $c^{\delta\lambda}$, $c^{\bar\sigma'}$, $c^{\bar D'}$, and $c^{\bar\sigma^{(3)}}$ introduced in Eq.~\eqref{eq:scalings_asym}. After calculating $\varphi^m$ and $\hat\varphi^m$ up to order $m^3$, the Landau function is again obtained in the form of Eq.~\eqref{eq:landau_eq_asym}, with the only change being that $\delta\lambda$ is modified to
\begin{align}
\delta\lambda \equiv \lambda - \lambda_\cc - \frac{2\bar D}{\bar\sigma} \delta\rho
\end{align}
for $\lambda_\cc$ given by the unshifted form Eq.~\eqref{eq:lambda_c_eq}.

\begin{figure}
\includegraphics[width=0.49\textwidth]{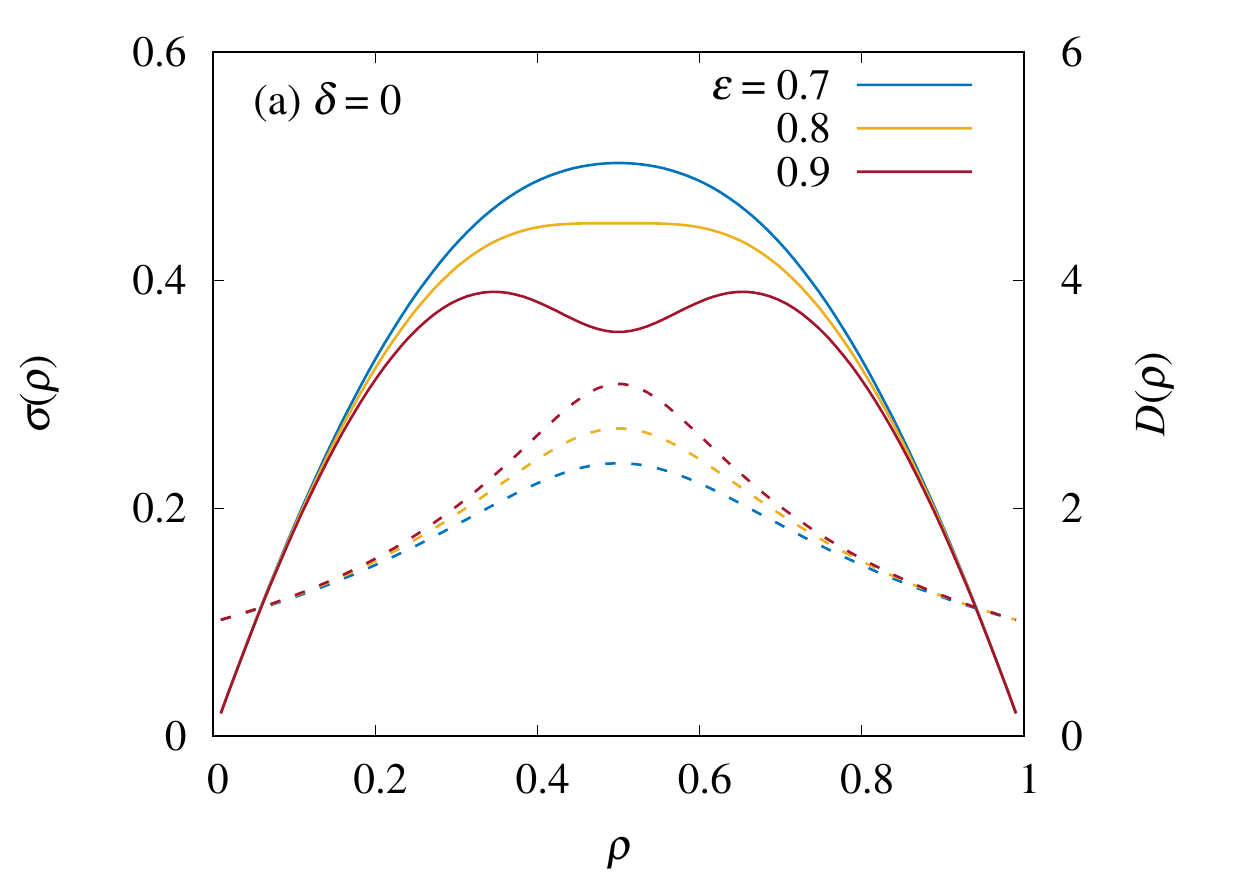}
\includegraphics[width=0.49\textwidth]{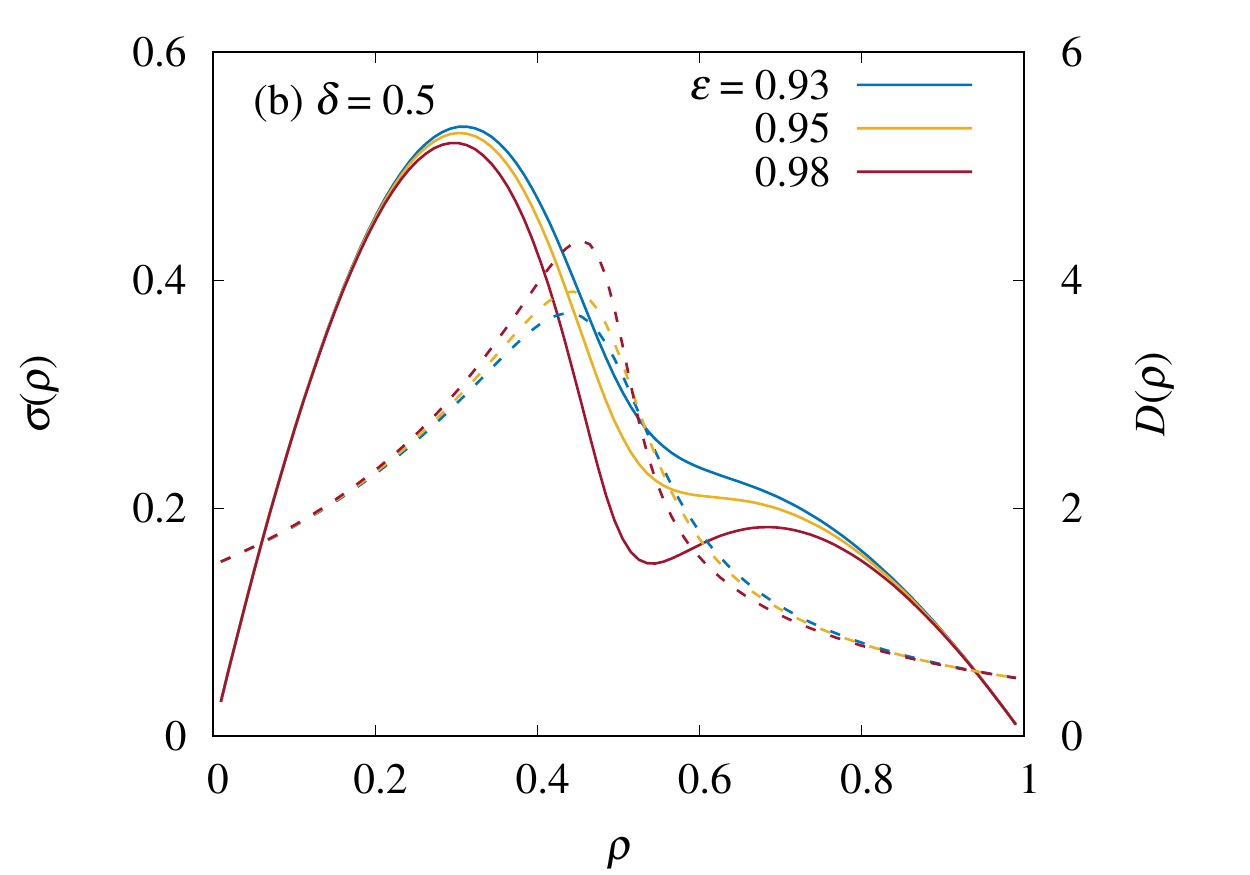}
\caption{\label{fig:kls} Density dependence of transport coefficients of the KLS model. The mobility coefficient $\sigma(\rho)$ is shown in solid lines, and the diffusion coefficient $D(\rho)$ in dashed lines for (a) particle--hole symmetric and (b) asymmetric systems.}
\end{figure}

\section{Transport coefficients of the Katz--Lebowitz--Spohn model} \label{sec:kls}

In the following we present explicit formulas for the transport coefficients of the Katz--Lebowitz--Spohn (KLS) model, which can be obtained by the methods of~\cite{hager_minimal_2001,Krapivsky_2013,krapivsky_unpub}. The diffusion coefficient is given by
\begin{align}
D(\rho) = \frac{\mathcal{J}(\rho)}{\chi(\rho)} \,,
\end{align}
where $\mathcal{J}(\rho)$ is the average current of the totally asymmetric version of the model satisfying
\begin{equation}
\label{current}
\mathcal{J}(\rho) = \frac{\nu [1+\delta(1-2\rho)]-\epsilon\sqrt{4\rho(1-\rho)}}{\nu^3} \,,
\end{equation}
and $\chi(\rho)$ is the compressibility given by
\begin{equation}
\label{compress:def}
\chi(\rho) = \rho(1-\rho)\sqrt{(2\rho-1)^2 + 4\rho(1-\rho)e^{-4\beta}} \,,
\end{equation}
with
\begin{equation}
\label{lambda:Krug}
\nu \equiv \frac{1+ \sqrt{(2\rho-1)^2 + 4\rho(1-\rho)e^{-4\beta}}}{\sqrt{4\rho(1-\rho)}} \,,
\quad e^{4\beta} \equiv \frac{1+\epsilon}{1-\epsilon} \,.
\end{equation}
Then the mobility coefficient $\sigma(\rho)$ is obtained from the Einstein relation
\begin{align}
\sigma(\rho) = 2 D(\rho) \, \chi(\rho) \,.
\end{align}

Behaviors of the transport coefficients obtained from the above results are illustrated in Fig.~\ref{fig:kls}. When the system has a full particle--hole symmetry ($\delta = 0$), $\sigma(\rho)$ has a local extremum at $\rho = 1/2$, which becomes a local minimum for sufficiently strong repulsion ($\epsilon > 4/5$), as shown in Fig.~\ref{fig:kls}(a). In the absence of the symmetry ($\delta \neq 0$), $\sigma(\rho)$ has a local extremum at a different value of $\rho$, which again becomes a local minimum for sufficiently large $\epsilon$ (see Fig.~\ref{fig:kls}(b)).

\section{Saddle point stability in the presence of bulk field} \label{sec:stability_field}

An argument proposed by~\cite{Shpielberg2016} states that a time-independent saddle-point solution $\rho_0(x)$ satisfying
\begin{align} \label{eq:shpielberg_condition}
	D'(\rho_0(x))\sigma'(\rho_0(x)) \ge D(\rho_0(x))\sigma''(\rho_0(x))
\end{align}
across the system ($0 \le x \le 1$) is stable regardless of the bulk field $E$. If true, the argument forbids symmetry-breaking transitions in bulk-driven systems with a local maximum of $\sigma(\rho)$ (e.g.~in the WASEP), which are predicted by our study. In the following we address this apparent contradiction.

As stated in Eq.~\eqref{eq:psi_least_action}, the scaled CGF is obtained from minimization of the action
\begin{align}
	S \equiv \int_0^T \mathrm{d}t \, \int_0^1 \mathrm{d}x \,
	\left[\hat\rho\dot\rho - H(\rho,\hat\rho)\right].
\end{align}
Suppose that a time-independent saddle-point solution is given by $\rho(x,t) = \rho_0(x)$ and $\hat\rho(x,t) = \hat\rho_0(x)$. For this solution to be unstable, in the vicinity there must be another saddle-point solution $\rho(x,t) = \rho_0(x) + \varphi(x,t)$ and $\hat\rho(x,t) = \hat\rho_0(x) + \hat\varphi(x,t)$ whose value of $S$ is smaller. In~\cite{Shpielberg2016} it was shown that the change of action due to $\varphi(x,t)$ and $\hat\varphi(x,t)$ is given by
\begin{align}
	\Delta S \simeq \int_0^T \mathrm{d}t \, \int_0^1 \mathrm{d}x \,
	\left[\frac{D'(\rho_0)\sigma'(\rho_0)-D(\rho_0)\sigma''(\rho_0)}{4D(\rho_0)}\left(\nabla\hat\rho_0\right)\left(\nabla\hat\rho_0+2E\right)\varphi^2 + \frac{\sigma(\rho_0)}{2}\left(\nabla\hat\varphi\right)^2\right]
\end{align}
up to the leading-order contributions. In the absence of the bulk field ($E = 0$), it is clear that Eq.~\eqref{eq:shpielberg_condition} implies $\Delta S \ge 0$, implying the stability of $\rho_0(x)$ and $\hat\rho_0(x)$. If $E \neq 0$, the sign of $\left(\nabla\hat\rho_0\right)\left(\nabla\hat\rho_0+2E\right)$ determines whether Eq.~\eqref{eq:shpielberg_condition} remains the sufficient condition for stability.

Defining $u \equiv \nabla\hat\rho_0 + E$, from the second equation of Eq.~\eqref{eq:saddle_traj_time_indept} one finds~\cite{Shpielberg2016}:
\begin{align}
	\frac{\nabla u}{u^2 - E^2} = -\frac{\sigma'(\rho_0)}{2D(\rho_0)}\, .
\end{align}
The l.h.s.~of this equation can be written as
\begin{align}
	\frac{\nabla u}{u^2 - E^2} = \begin{cases}
 	-\frac{1}{E}\nabla \mathrm{arctanh} \frac{u}{E} &\text{ if $|u| < |E|$,} \\
 	-\frac{1}{E}\nabla \mathrm{arccoth} \frac{u}{E} &\text{ if $|u| > |E|$.}
 \end{cases}
 \label{eq:diff_u}
\end{align}
In~\cite{Shpielberg2016}, only the latter case is considered, so that one can write $u = E \coth (Eh)$ where $h$ satisfies $\nabla h = \sigma'(\rho)/\left[2D(\rho)\right]$. Then we obtain
\begin{align}
	\left(\nabla\hat\rho_0\right)\left(\nabla\hat\rho_0+2E\right)
	= u^2 - E^2 = \frac{E^2}{\sinh^2 (Eh)} > 0 \,,
\end{align}
which ensures that Eq.~\eqref{eq:shpielberg_condition} is still a sufficient condition for the stability of $\rho_0(x)$ and $\hat\rho_0(x)$.

However, close to the symmetry-breaking transition points of the WASEP, one can show that $|u| < |E|$ is satisfied across the system. In this case, the first case of Eq.~\eqref{eq:diff_u} should be used. This implies $u = E \tanh (Eh)$, from which we obtain
\begin{align}
	\left(\nabla\hat\rho_0\right)\left(\nabla\hat\rho_0+2E\right)
	= u^2 - E^2 = -\frac{E^2}{\cosh^2 (Eh)} < 0 \,.
\end{align}
Since the sign of $\left(\nabla\hat\rho_0\right)\left(\nabla\hat\rho_0+2E\right)$ is inverted, Eq.~\eqref{eq:shpielberg_condition} is no longer a sufficient condition for the stability of $\rho_0(x)$ and $\hat\rho_0(x)$. Thus, while Eq.~\eqref{eq:shpielberg_condition} gives the correct sufficient condition for stability in the absence of the bulk field $E$, it does not apply to the case when $\bar\sigma'' < 0$ and $E \neq 0$.

\bibliography{symmetry_breaking_BKL}

\end{document}